\documentclass[amsmath,amssymb, aps, pra,reprint]{revtex4-1}

\usepackage{float}
\usepackage{amsmath,amssymb}
\usepackage[english]{babel}
\usepackage{upgreek}
\usepackage{dsfont}
\usepackage{graphicx}
\usepackage{graphicx}
\usepackage{color}

\DeclareMathOperator{\Tr}{tr}
\newcommand{\op}[1]{{\sf #1}}

\newcommand{\cD}{{\mathcal D}}

\newcommand{\cO}{{\mathcal O}}

\newcommand{\ZZ}{{\mathbb Z}}

\newcommand{\oH}{{\mathsf H}}

\newcommand{\eps}{\varepsilon}

\newcommand{\ket}[1]{\left \vert #1 \right \rangle}
\newcommand{\braket}[2]{\langle #1 \vert #2 \rangle}
\newcommand{\matel}[3]{{\left\langle \vphantom{#1 #2 #3} #1 \,\right\vert
\left.
 \hspace{-0.15em} \vphantom{#1 #2 #3} #2 \,\right\vert \left.
 \hspace{-0.15em} \vphantom{#1 #2 #3} #3\right\rangle}}
\newcommand{\ketbra}[2]{\vert #1 \rangle \langle #2 \vert}
\newcommand{\be}{\begin{equation}}
\newcommand{\ee}{\end{equation}}
\newcommand{\bea}{\begin{eqnarray}}
\newcommand{\eea}{\end{eqnarray}}

\newcommand{\mitl}[1]{\left \langle #1 \right \rangle}
\newcommand{\un}{{\mathds 1}}

\newcommand{\oJ}{\textsf{\textbf{J}}}
\newcommand{\orcm}{\textsf{\textbf{r}}}
\newcommand{\opcm}{\textsf{\textbf{p}}}
\newcommand{\oOmega}{\op{\Omega}}
\newcommand{\oL}{\textsf{\textbf{L}}}
\newcommand{\oK}{\textsf{\textbf{K}}}
\newcommand{\rD}{\mathrm{D}}
\newcommand{\rI}{\mathrm{I}}
\newcommand{\rR}{\mathrm{R}}

\newcommand{\rGamma}{\mathrm{\Gamma}}

\newcommand{\iGamma}{{\it \Gamma}}

\begin{document}
\title{Rotational friction and diffusion of quantum rotors}

\author{Benjamin A. Stickler}
\author{Bj\"{o}rn Schrinski}
\author{Klaus Hornberger}
\affiliation{
 University of Duisburg-Essen, Faculty of Physics, Lotharstra\ss e 1, 47048 Duisburg, Germany}

\begin{abstract}
We present the  Markovian quantum master equation describing rotational decoherence, friction, diffusion, and thermalization of planar, linear, and asymmetric rotors in contact with a thermal environment. It describes how an arbitrary initial rotation state decoheres and evolves toward a Gibbs-like thermal ensemble, as we illustrate numerically for the linear and the planar top, and it yields the expected rotational Fokker-Planck equation of Brownian motion in the  semiclassical limit. 
\end{abstract}

\maketitle

\textit{Introduction ---}  A quantum point particle moving in a thermal bath is subject to  random interactions with the environmental degrees of freedom. They affect the particle in two ways: (i) initial superpositions of different positions quickly decohere and (ii) the particle gradually thermalizes with its surroundings. In many situations, the associated dynamics is well modeled by the Markovian master equation of quantum Brownian motion \cite{caldeira1983,breuer2002}. But what if the particle is not point-like and hence able to rotate? How can the resulting rotational decoherence, friction and diffusion be described quantum mechanically?

Beyond its conceptual significance, this question becomes increasingly relevant for state-of-the-art experiments.
Numerous experimental studies demonstrate rotational manipulation and control of molecules  \cite{stapelfeldt03,holmegaard2009,korobenko2014,benko2015,tenney2016,shepperson2017} and recently also of nanoparticles \cite{kane2010,arita2013,kuhn2015,hoang2016,coppock2016,kuhn2017a,kuhn2017b,delord2017}. Cooling the rotation state into the quantum regime was successfully implemented for small molecules \cite{ni2008,danzl2010,staanum2010,zeppenfeld2012,stuhl2012,lien2014,barry2014,gloeckner2015,prehn2016,bohn2017}, and is in reach for nano- to micrometer-sized objects \cite{stickler2016a,hoang2016,zhong2017}. Conceivable applications include  orientation-dependent metrology 
\cite{kraus2015,deiss2014,schoun2017}, ultra-cold chemistry \cite{bell2009,ospelkaus2010b,moses2017}, highly-sensitive torque sensors \cite{hoang2016,kuhn2017b}, realizations of a quantum heat engine \cite{roulet2017}, levitated nanomagnets \cite{rusconi2016}, dissipative dynamics of angulons \cite{schmidt2015}, tests of objective collapse models \cite{schrinski2017} and orientational quantum revivals \cite{seideman1999,stickler2018b}. The interpretation of such experiments will rely heavily on a theoretical assessment of the rotor-dynamics in presence of an environment. 

Here, we present the general Markovian quantum master equation describing rotational friction, diffusion, and thermalization of rigid rotors. It is the natural generalization of the master equation of Brownian center-of-mass motion \cite{caldeira1983,breuer2002,vacchini2009}, valid if the bath is sufficiently dilute or its temperature is high enough to warrant a Markovian description. Unlike the center-of-mass momentum, however, the angular momentum components do not commute, implying that the orientational degrees of freedom cannot be decoupled, and the friction and diffusion tensors depend necessarily on the particle orientation expressed in terms of rotation matrices. These characteristics of orientation and rotation render their quantum dynamics substantially different from the center-of-mass motion, so that the presented master equation is not a straight-forward extension thereof. As in the classical theory \cite{lukas}, it will turn out to be pertinent to use a coordinate-independent formulation in terms of rotation matrices and angular momentum vectors rather than the canonical phase space variables. The presented master equation reduces to pure orientational decoherence \cite{stickler2016b,zhong2016,papendell2017} in the high-temperature limit and to the expected Fokker-Planck equation \cite{lukas} in the semiclassical limit.

\begin{figure}[t]
  \centering
  \includegraphics[width=\columnwidth]{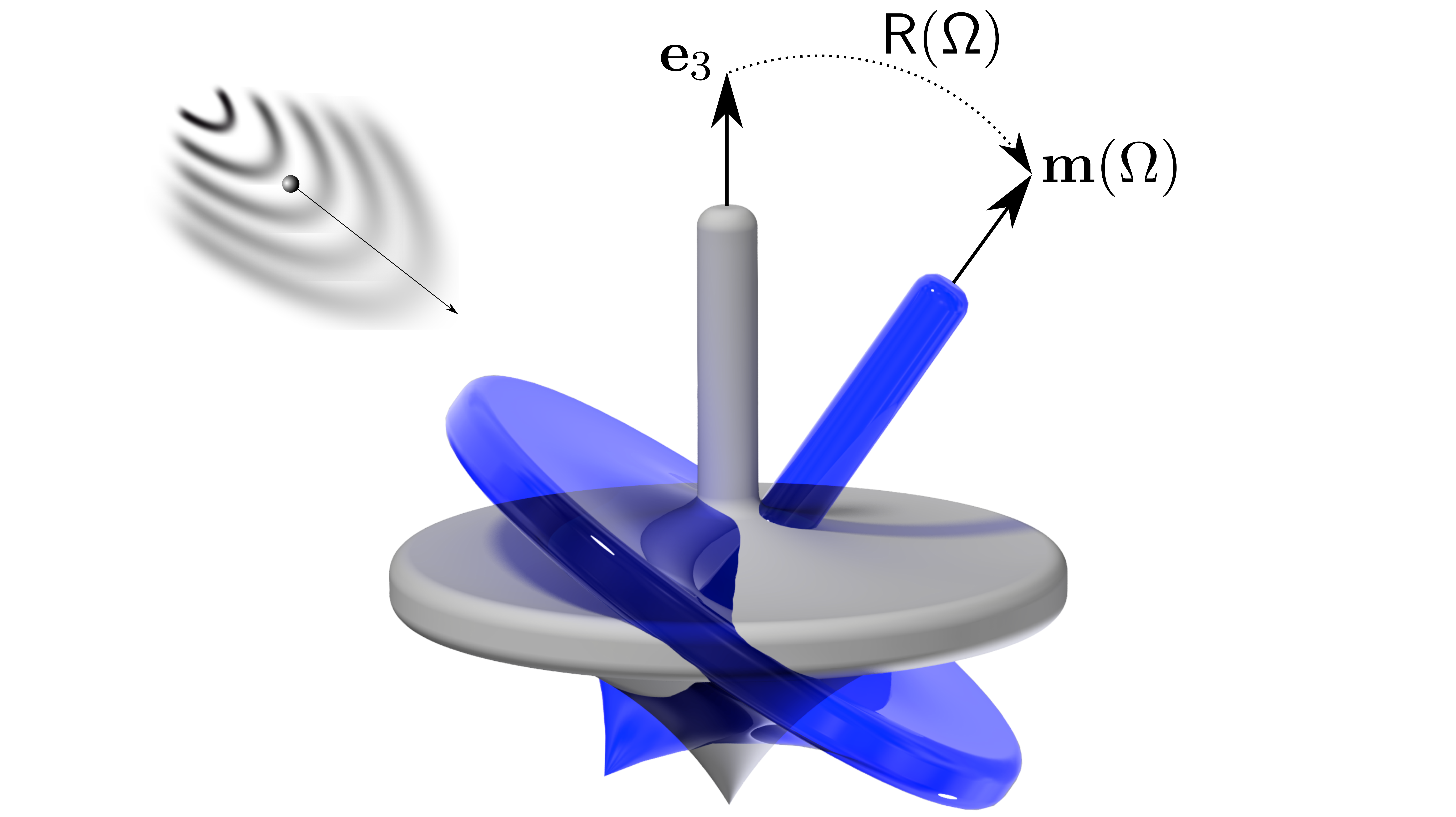}
  \caption{A rigid rotor immersed in a thermal environment receives random angular momentum kicks analogous to the momentum kicks experienced by a Brownian particle. These interactions decohere an initial superposition of two different orientations and gradually thermalize the rotation state.}
\label{fig:Springer}
\end{figure}

{\it Classical Thermalization --- } It is useful to briefly review the classical description of rotational thermalization of a rigid body of orientation $\Omega$ (parametrized e.g.\ by the Euler angles) and angular momentum ${\bf J}$. In absence of an external torque, environment-induced friction and diffusion can be described by the stochastic differential equation ${\rm d}{\bf J} = -\rGamma(\Omega) {\bf J}\mathrm{d}t + \mathrm{d}{\bf N}_t$. The first term accounts for rotational friction with the orientation-dependent friction tensor $\rGamma(\Omega) = \rR(\Omega)\rGamma_0\rR^{\rm T}(\Omega)$. Here, the orthogonal matrix $\rR(\Omega)$ serves to rotate the particle from a reference orientation $\Omega = 0$ to the current one, see Fig.~\ref{fig:Springer}, and $\rGamma_0$ is the friction tensor at $\Omega = 0$. The strength and direction of the random angular momentum kicks ${\rm d}{\bf N}_t = \sqrt{2 \rD(\Omega)} \mathrm{d} {\bf W}_t$ is determined by the diffusion tensor $\rD(\Omega)$ \cite{papendell2017}, while ${\rm d}{\bf W}_t$ is a vector Wiener process. The angular momentum dynamics are complemented by the equation of motion for particle orientation, ${\rm d}{\rR}(\Omega) = \rI^{-1}(\Omega) {\bf J} \times \rR(\Omega) {\rm d}t$. Here, we introduced the tensor of inertia $\rI(\Omega)$ whose eigenvalues $I_i$ are the moments of inertia. These two stochastic equations determine the Brownian rotation dynamics of an arbitrary particle.

The stochastic motion can be equivalently described by the deterministic evolution of the probability density $h_t(\Omega,{\bf J})$ \cite{gardiner1985,vankampen1995,risken1996}. It contains both the free rotational dynamics and a non-conservative part accounting for the interaction with the environment, $\partial_t h_t = \partial^{\rm rot}_t h_t + \partial_t^{\rm nc} h_t$. While the first part is determined by the Hamilton function $H= {{\bf J} \cdot \rI^{-1}(\Omega) {\bf J}}/2$, the second part takes the form of a Fokker-Planck equation,
\begin{equation} \label{eq:fp}
\partial_t^{\rm nc} h_t(\Omega,{\bf J}) = \nabla_{{\bf J}}\cdot[\rGamma(\Omega){\bf J}h_t(\Omega,{\bf J})]  + \nabla_{{\bf J}}\cdot \rD(\Omega)\nabla_{{\bf J}} h_t(\Omega,{\bf J}).
\end{equation}
The dynamics of the first two moments of ${\bf J}$ follow directly as
\begin{subequations}\label{eq:Lmoments}
	\begin{align}
	\partial_t \langle{\bf J}\rangle = & -\langle\rGamma(\Omega){\bf J} \rangle, \\
	\partial_t \langle {\bf J}^2 \rangle = & - 2 \langle {\bf J} \cdot \rGamma(\Omega) {\bf J} \rangle + 2 \langle {\rm Tr}[\rD(\Omega)] \rangle, \\
	\partial_t\langle{\bf J}\otimes{\bf J}\rangle = & -\langle\rGamma(\Omega)\,{\bf J}\otimes{\bf J} + {\bf J}\otimes{\bf J}\,\rGamma^{\rm T}(\Omega)\rangle + 2\langle\rD(\Omega)\rangle.
	\end{align}
\end{subequations}
As expected for Brownian motion, friction reduces the mean (angular) momentum (\ref{eq:Lmoments}a), while diffusion increases its variance and covariance with a constant rate determined by (\ref{eq:Lmoments}b) and (\ref{eq:Lmoments}c). [$\mathrm{Tr}(\cdot)$ refers to the matrix trace, as opposed to the operator trace $\mathrm{tr}(\cdot)$ used below.]

Using the fluctuation-dissipation relation $\rD(\Omega) = k_{\rm B} T \rGamma(\Omega) \rI(\Omega)$ in Eq.~\eqref{eq:fp}, one finds that the rotor thermalizes toward the Gibbs state $\exp( - H/ k_{\rm B} T)/Z$ with mean energy
\begin{equation} \label{eq:equien}
\langle H \rangle = \frac{1}{2} \mitl{{\bf J} \cdot \rI^{-1}(\Omega) {\bf J}} = \frac{f}{2} k_{\rm B} T,
\end{equation}
where $f = {\rm rank}[\rI(\Omega)]$ is the number of rotational degrees of freedom. For a given particle shape, the rotational friction tensor can be calculated microscopically from kinetic gas theory \cite{dahneke1973,eisner1981,lukas}. The Fokker-Planck description \eqref{eq:fp} allows general statements about thermalization \cite{risken1996}, and is best suited for comparison with the quantized rotation dynamics.

Quantum-classical consistency demands that the quantum master equation of rotational friction and diffusion describes the same dynamics as Eq.~\eqref{eq:fp} in the semiclassical limit. This means that the equations for the first and second moments of the angular momentum operator $\oJ$ (operators are denoted by {\it sans-serif} characters) must coincide with their classical equivalents \eqref{eq:Lmoments} up to corrections of order $\hbar$. Further, the steady state of the quantum master equation must approach the Gibbs state for large temperatures,
\begin{equation} \label{eq:stgibbs}
\rho_{\rm eq} = \frac{1}{Z}e^{- \oH/ k_{\rm B} T} + \cO(\hbar), \quad \text{with}\quad Z = \mathrm{tr} \left (e^{- \oH/ k_{\rm B} T} \right ),
\end{equation}
implying the equipartition of energies \eqref{eq:equien} to lowest order in $\hbar$.

\begin{figure*}
  \centering
  \includegraphics[width=0.99\textwidth]{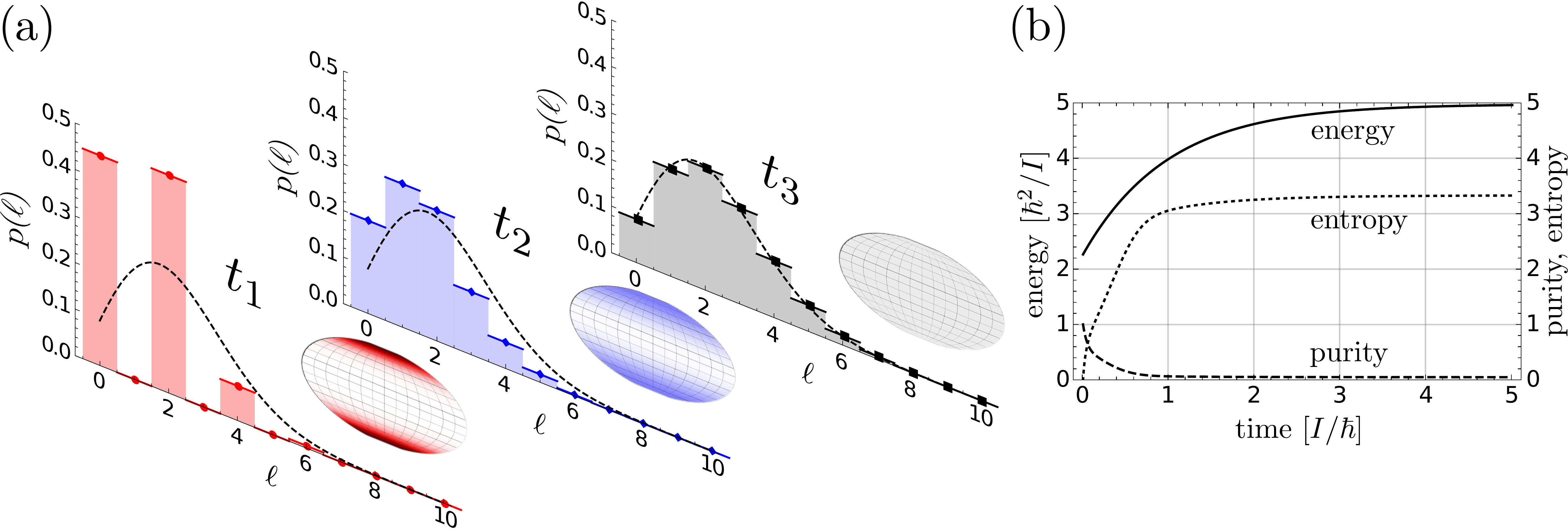}
  \caption{Time evolution of the linear rigid rotor with free Hamiltonian $\oH = \oJ^2/2 I$ and  dissipator \eqref{eq:dissnLB} for an initial superposition of pointing up- and downwards. (a) The histograms show the probability  $p_\ell = \sum_{m = -\ell}^\ell \langle \ell m \vert \rho \vert\ell m\rangle$ of observing the total angular momentum $\ell$ for $t_1 = 0$, $t_2=0.5 I/\hbar$, and $t_3=5 I/\hbar$. The dashed line represents the Gibbs state, and the insets display polar density plots of the orientational distribution $\langle \Omega \vert \rho\vert \Omega\rangle$ (Mollweide projection).  (b) Time dependence of the energy expectation value $\mitl{\oH}$ (solid line), the von Neumann entropy $-\mathrm{tr}(\rho \log \rho)$  (dotted line), and the purity $\mathrm{tr}(\rho^2)$ (dashed line).
  We use $\xi = 5$, $\iGamma = \hbar / I$, and $\sigma = 0.4$.}
\label{fig:linrot}
\end{figure*}

\textit{General Master Equation --- } We now establish the quantum master equation $\partial_t \rho = - i[ \oH, \rho]/\hbar + \cD \rho$ describing rotational friction and diffusion through the dissipator $\cD$. The latter can be heuristically derived from the Caldeira-Leggett equation \cite{caldeira1983,breuer2002} for $N$ rigidly connected point particles. Denoting the position operators of the individual point particles by $\orcm_n$ and the momentum operators by $\opcm_n$, the master equation reads
\begin{align} \label{eq:cd}
\partial_t \rho = & - \frac{i}{\hbar} \left [\oH + \frac{1}{2} \sum_{n = 1}^N \gamma_n  \left ( \orcm_n \cdot \opcm_n + (\orcm_n \cdot \opcm_n)^\dagger \right ) , \rho \right ]  \nonumber \\
& + \frac{2 k_{\rm B} T}{\hbar^2}  \sum_{n = 1}^N m_n \gamma_n\left [ \oL_n \cdot \rho \oL_n^\dagger - \frac{1}{2} \left \{ \oL^\dagger_n \cdot \oL_n,\rho \right \} \right ],
\end{align}
with $\oL_n = \orcm_n + i \hbar \opcm_n/4 m_n k_{\rm B}T$ and individual damping rates $\gamma_n>0$.

In order to account for the fact that all point particles are rigidly connected we introduce a quantum version of the rigid body approximation \cite{goldstein}. Denoting the orientation operator as $\oOmega$, it takes the center-of-mass to be at rest, $\orcm_n = \rR(\oOmega) {\bf r}_n^{(0)}$, and replaces the momenta $\opcm_n$ by operator-valued classical expressions for the velocity of the $n$-th particle multiplied by its mass, $-m_n \rR(\oOmega) {\bf r}_n^{(0)} \times \rR(\oOmega)\rI_0^{-1}\rR^{\rm T}(\oOmega)\oJ$, with ${\bf r}_n^{(0)}$ the $n$-th particle position and $\rI_0$ the tensor of inertia at orientation $\Omega = 0$. While the new momenta are non-hermitian, they ensure both  that $\oJ = \sum_n \orcm_n \times \opcm_n$ still holds after the replacements and that the energy renormalization in \eqref{eq:cd} vanishes. Notwithstanding the heuristic nature of this quantum rigid body approximation we will see that the resulting master equation has all desired properties.

Using the quantum rigid body approximation in \eqref{eq:cd} yields
\begin{align}
\partial_t \rho = & - \frac{i}{\hbar} \left [\oH, \rho \right ] + \frac{2 k_{\rm B} T}{\hbar^2} \sum_{n = 1}^N m_n \gamma_n \Biggl [ \rR(\oOmega) \oK_n \cdot \rho \oK_n^\dagger \rR^{\rm T}(\oOmega) \nonumber \\
& - \frac{1}{2} \left \{ \oK_n^\dagger \cdot\oK_n, \rho \right \} \Biggr ],
\end{align}
where $\oK_n = {\bf r}_n^{(0)} + i \hbar \rI_0^{-1} \rR^{\rm T}(\Omega) \oJ \times {\bf r}_n^{(0)}/4 k_{\rm B} T$. Subsuming the sum into the positive tensor
\begin{equation}
\label{eq:A0}
\widetilde{\rD}_0 = k_{\rm B} T \sum_{n = 1}^N m_n \gamma_n {\bf r}_n^{(0)} \otimes {\bf r}_n^{(0)} \equiv \sum_{k = 1}^3 \widetilde{D}_k {\bf d}_k^{(0)} \otimes {\bf d}_k^{(0)},
\end{equation}
one obtains the master equation
\begin{subequations} \label{eq:asi}
\begin{equation}
\cD \rho =  \sum_{k = 1}^3 \frac{2 \widetilde{D}_k}{\hbar^2} \left [ \textsf{\textbf{A}}_k \cdot \rho \textsf{\textbf{A}}^{\dagger}_k -\frac{1}{2}\left\{ \textsf{\textbf{A}}^{\dagger}_k\cdot\textsf{\textbf{A}}_k,\rho\right\}\right].
\end{equation}
It involves scalar products of the vectorial Lindblad operators
\begin{equation} \label{eq:LBasi}
\textsf{\textbf{A}}_k = {\bf d}_k(\oOmega) - \frac{i\hbar}{4 k_{\rm B} T}{\bf d}_k(\oOmega)\times\rI^{-1}(\oOmega)\oJ,
\end{equation}
\end{subequations}
with ${\bf d}_{k}(\Omega) = \rR(\Omega){\bf d}_k^{(0)}$, where ${\bf d}_k^{(0)}$ are the three normalized orthogonal eigenvectors of $\widetilde{\rD}_0$. Equation \eqref{eq:difftens} below shows that the ${\bf d}_k(\Omega)$ are the eigenvectors of the diffusion tensor whose eigenvalues $D_k$ fix the $\widetilde{D}_k$.

Equations \eqref{eq:asi} specify the quantum Brownian rotation dynamics expected for asymmetric rigid tops. They give rise to the moment dynamics \eqref{eq:Lmoments} to leading order in $\hbar$, and they ensure that $\rho$ approaches the steady state \eqref{eq:stgibbs} with energy expectation value \eqref{eq:equien} for small $\hbar^2/k_{\rm B} T I_{\rm min}$, with $I_{\rm min}$ the minimal moment of inertia. All this can be checked by straight-forward but lengthy calculations taking into account that the $\textsf{\textbf{A}}_k$ and their components do not commute, as explained in \cite{suppinfo}.

While the first term of the Lindblad operators \eqref{eq:LBasi} represents the particle orientation through ${\bf d}_k(\Omega)$, the second is proportional to the quantized (but not hermitized) rate of change, $\dot{\bf d}_k(\Omega) = \rI^{-1}(\Omega){\bf J} \times {\bf d}_k(\Omega)$. Equation \eqref{eq:LBasi} is thus the rotational analogue of the Lindblad operator $\op{L}=\op{x}+i\hbar\op{p}/4 m k_{\rm B} T$ of one-dimensional center-of-mass thermalization in quantum Brownian motion \cite{caldeira1983,breuer2002,wiseman2009,vacchini2009}. In contrast to the latter, the three Lindblad operators \eqref{eq:LBasi} do not commute, accounting for the facts that the three principal axis of a rigid rotor cannot be rotated independently and that the components of the angular momentum vector do not commute.

The vectors ${\bf d}_k(\Omega)$ are the eigenvectors of the diffusion tensor $\rD(\Omega)$, as can be demonstrated by calculating the second moments \eqref{eq:Lmoments} using Eq.~\eqref{eq:asi}. Comparison with \eqref{eq:Lmoments} shows that
\begin{equation} \label{eq:difftens}
\rD(\Omega) = \sum_{k = 1}^3 \widetilde{D}_k \left [\un - {\bf d}_k(\Omega) \otimes {\bf d}_k(\Omega) \right ],
\end{equation}
with eigenvalues $D_k = \widetilde{D}_i + \widetilde{D}_j$, so that $\widetilde{D}_k = (D_i + D_j - D_k)/2$, where $(i,j,k)$ are permutations of $(1,2,3)$. This relation implies that the master equation \eqref{eq:asi} is completely positive ($\widetilde{D}_k \geq 0$) only if $D_i + D_j \geq D_k$ (even though the localization rate is always positive). The same inequality is implied by the corresponding classical derivation of Brownian motion \cite{suppinfo}, where a more general diffusion tensor can be obtained if the diffusion of the individual point particles is not isotropic. It remains an open question how to extend this to the quantum regime.

The semiclassical limit of \eqref{eq:asi}  gives the rotational Fokker-Planck equation \eqref{eq:fp} with diffusion tensor \eqref{eq:difftens} and friction tensor $\rGamma(\Omega) =  \rD(\Omega) \rI^{-1}(\Omega) / k_{\rm B} T$. This can be shown by adopting the treatment in Ref.~\cite{papendell2017}, i.e.\ first expressing \eqref{eq:asi} in the quantum phase space of the orientation state \cite{fischer2013,gneiting2013}, approximating the discrete angular momentum quantum numbers by continuous variables, and then evaluating the limit $\hbar \to 0$.

Another limiting case is that the rotor is tightly aligned by an external potential, so that its dynamics are librational rather than rotational. If the angle coordinates can then be approximated harmonically, a linearization of the rotation matrix in the angle operators yields Lindblad operators reminiscent of center-of-mass Brownian motion with positions and momenta replaced by angles and their canonically conjugate momenta.

In what follows we will specialize the master equation \eqref{eq:asi}, which is valid for general rotors, to the cases of the linear and planar rigid tops and illustrate their thermalization dynamics.

{\it Linear rotors ---} The orientation of a linear rigid rotor is specified by the direction of its symmetry axis ${\bf m}(\Omega)$, so that $\rI(\Omega) = I [ \un - {\bf m}(\Omega) \otimes {\bf m}(\Omega)]$. Accordingly, friction and diffusion orthogonal to the symmetry axis, for instance due to specular gas scattering \cite{lukas}, are described by the tensors  $\rGamma(\Omega) = \iGamma[ \un - {\bf m}(\Omega) \otimes {\bf m}(\Omega)]$ and $\rD(\Omega) = D[ \un - {\bf m}(\Omega) \otimes {\bf m}(\Omega)]$ with $D=k_{\rm B} T \iGamma I$. This implies that one eigenvalue of the diffusion tensor is zero, while the two eigenvalues associated with the two directions perpendicular to ${\bf d}_1(\Omega) = {\bf m}(\Omega)$ are $D$.

Calculating $\widetilde{D}_k$ according to \eqref{eq:difftens} yields the dissipator
\begin{subequations} \label{eq:dissnLB}
\begin{align}
\mathcal{D}\rho= & \frac{2 D}{\hbar^2}
\left[ \textsf{\textbf{A}} \cdot \rho \textsf{\textbf{A}}^{\dagger} -\frac{1}{2}\left\{ \textsf{\textbf{A}}^{\dagger}\cdot\textsf{\textbf{A}},\rho\right\}\right],
\label{eq:LBTFull}
\end{align}
with the vectorial Lindblad operators
\begin{align} \label{eq:LB}
\textsf{\textbf{A}} = {\bf m}(\oOmega) - \frac{i\hbar}{4 k_{\rm B} T I}{\bf m}(\oOmega)\times\oJ.
\end{align}
\end{subequations}

Inserting \eqref{eq:LB} into \eqref{eq:LBTFull} yields
\begin{eqnarray}  \label{eq:rotcalleg}
\cD \rho  & = & - \frac{i \iGamma}{2 \hbar}  \left [{\bf m}(\oOmega) \times \oJ \cdot \rho {\bf m}(\oOmega) + {\bf m}(\oOmega)  \cdot \rho \oJ \times  {\bf m}(\oOmega) \right ]\nonumber\\
&& \frac{2 D}{\hbar^2} \left [ {\bf m}(\oOmega) \cdot \rho {\bf m}(\oOmega) - \rho \right ]  +\cO\left(\frac{\hbar^2}{k_{\rm B}T I}\right)\,.
\end{eqnarray}
Here, the first term is independent of temperature, linear in $\oJ$ and describes rotational friction.
The second term is linear in $T$ (since $D=k_{\rm B}T\iGamma I$) and describes angular momentum diffusion as well as an exponential decay of the orientational coherences $\matel{\Omega}{\rho}{\Omega'}$ with the rate
\begin{equation}\label{eq:locrate}
F(\Omega,\Omega') = \frac{2 D}{\hbar^2} \left [1 - {\bf m}(\Omega)\cdot {\bf m}(\Omega') \right ].
\end{equation}
The remaining terms in \eqref{eq:rotcalleg} are proportional to $1/T$, quadratic in $\op{J}$ and ensure complete positivity; like in the center-of-mass case, they can be neglected for sufficiently large temperatures. The special case that a symmetry in the environmental interaction prevents the complete localization 
\eqref{eq:locrate} can also be accounted for, as discussed in the supplements \cite{suppinfo}.

In order to study the thermalization dynamics described by \eqref{eq:dissnLB}, we solve the master equation numerically  with the free Hamiltonian $\oH = \oJ^2 / 2 I$ and calculate analytically the corresponding steady state $\rho_{\rm eq}$. The latter can be determined by noting that the equation $-i[\oH,\rho_{\rm eq}] /\hbar + \cD\rho_{\rm eq} = 0$ implies that $\rho_{\rm eq}$ is diagonal in the angular momentum basis, $\rho_{\rm eq} = \sum_{\ell m} \rho_{\rm eq}^{\ell m} \ketbra{\ell m}{\ell m}$. Then, the unitary part vanishes and $\cD \rho_{\rm eq} = 0$ yields a set of coupled equations for the coefficients $ \rho_{\rm eq}^{\ell m}$. It can be solved explicitly \cite{suppinfo},
\begin{align} \label{eq:linrotstat}
 \rho_{\rm eq}^{\ell m} =\frac{1}{Z}
\binom{2\xi}{\ell}^2\binom{2\xi+\ell+1}{\ell}^{-2} 
\end{align}
in terms of $\xi =  2 I k_{\rm B}T/\hbar^2$. The steady state approaches the Gibbs state \eqref{eq:stgibbs} for large temperatures, $\rho_{\rm eq}^{\ell m} \sim \exp [ - \ell (\ell+1)/\xi]/Z$ as $\xi \to\infty$, as can be checked using Stirling's formula. From the existence of the steady state \eqref{eq:linrotstat} it follows that the relative entropy $S(\rho \| \rho_{\rm eq}) = - \mathrm{tr}[\rho(\log \rho - \log \rho_{\rm eq})] \leq 0$ increases monotonically with time and vanishes only for $\rho = \rho_{\rm eq}$ \cite{breuer2002}. Thus, an arbitrary initial state converges toward $\rho_{\rm eq}$.

\begin{figure*}
 \centering
 \includegraphics[width = 0.95\textwidth]{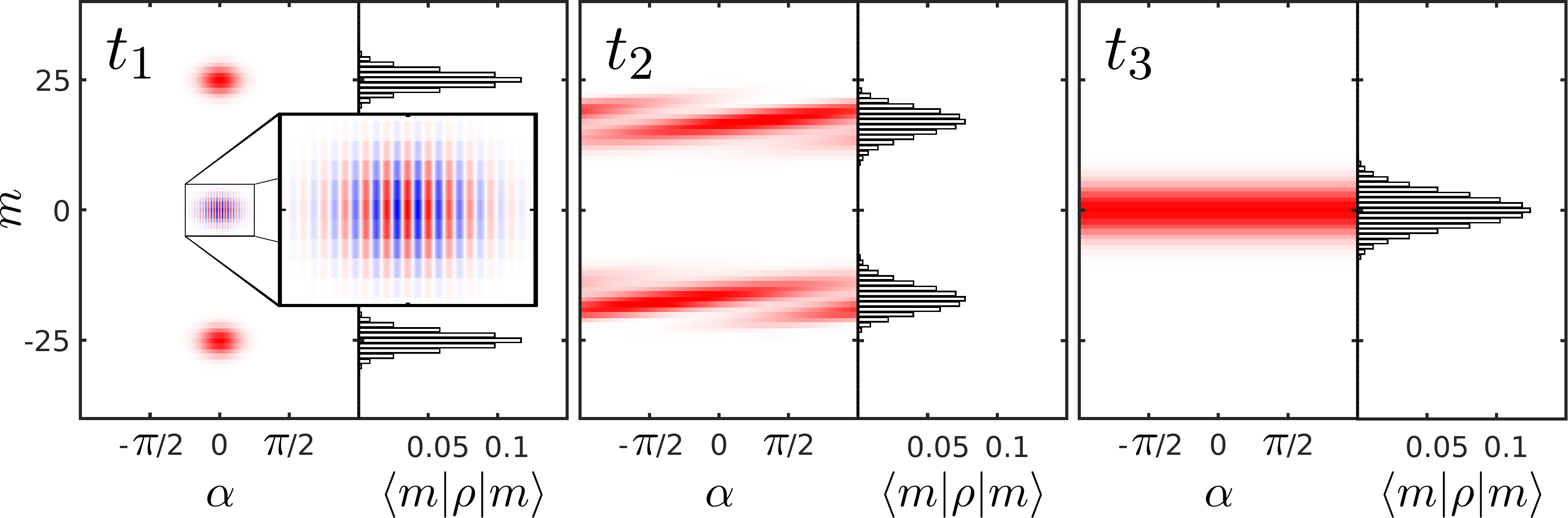}
 \caption{Time evolution of the planar rotor. The left sub-panels show the Wigner function $w_m(\alpha)$, and the right sub-panels the momentum marginals $\langle m \vert \rho \vert m \rangle$ for $t_1 = 0$, $t_2 = 4\pi I/ 10 \hbar$, and $t_3 = 4\pi I /\hbar$. The initial state is a superposition of two Gaussian states $\psi_0(\alpha)\propto\exp(i m_0\alpha + \cos \alpha/4\sigma^2)$  with $m_0 = \pm 25$ and $\sigma = 0.2$. Its coherence is indicated by the fringe structure (with negative values in blue), see inset. The plots are obtained by computing the Wigner function \cite{mukunda1979,rigas2011} of the time evolved density operator. The parameters are chosen as $\xi = 20$ and $\iGamma =  \hbar/\pi I$.} \label{fig:plrot}
\end{figure*}

We now  simulate numerically the dynamics for the pure initial rotor state $
\braket{\Omega}{\psi_0} \propto \exp\left(- \vert {\bf e}_z \times {\bf m}(\Omega) \vert^2/ 2\sigma^2 \right)$,
representing  a superposition of pointing upwards and downwards along the $z$-axis with width $\sigma$. Its time evolution is shown in Fig. \ref{fig:linrot}. The initial superposition first decoheres into a mixture of the up- and downwards orientation of the rotor, as evident from the purity. On the longer timescale $1/\iGamma$, the rotor approaches thermal equilibrium, as indicated by the energy expectation value and the von Neumann entropy. The final state, given by Eq. \eqref{eq:linrotstat}, is  already well approximated by the Gibbs state, even though the thermal occupation number $\overline{\ell}$, defined via $\overline{\ell}(\overline{\ell} + 1)  =  \xi$, is as low as $\overline{\ell} \simeq 2.7$.

\textit{Planar Rotors ---} If the rotor is confined to the $xy$-plane a single angle $\upalpha$ suffices to describe the orientation, ${\bf e}_r(\upalpha) = {\bf e}_x \cos \upalpha + {\bf e}_y \sin \upalpha$. The corresponding angular momentum operator points into the $z$-direction, $\oJ = {\bf e}_z \op{p}_\alpha$ and has discrete eigenvalues $\hbar m$, $m \in \ZZ$. The Lindblad operator takes on the form
\begin{equation} \label{eq:LBpl}
\textsf{\textbf{A}} = {\bf e}_r(\upalpha) + \frac{i\hbar}{4 k_{\rm B} T I} {\bf e}_\varphi(\upalpha) \op{p}_\alpha,
\end{equation}
where ${\bf e}_\varphi(\alpha) = {\bf e}_z \times {\bf e}_r(\alpha)$.

The action of the dissipator \eqref{eq:LBTFull} can be conveniently expressed in terms of the Wigner function $w_m(\alpha)$
\cite{mukunda1979,rigas2011}, 
\begin{align}
\partial_t^{\rm nc} w_{m}(\alpha) =& \frac{\iGamma}{2} \left [ (m+1)w_{m+1}(\alpha)-(m-1)w_{m-1}(\alpha) \right ]\nonumber\\
& + D \frac{w_{m+1}(\alpha) - 2w_{m}(\alpha) + w_{m-1}(\alpha)}{\hbar^2} 
.\label{eq:planrotTE}
\end{align}
As in Eq.~\eqref{eq:rotcalleg} we dropped the term proportional to $1/T$ (which vanishes in the semiclassical limit). Expression (\ref{eq:planrotTE}) is the discretized version of the Fokker-Planck equation, which is thus recovered in the semiclassical limit at fixed $p_\alpha=\hbar m$.

The stationary solution of the quantum planar rotor follows as $\rho_{\rm eq} = \sum_{m \in\ZZ} \rho_{\rm eq}^m \ketbra{m}{m}$ with \cite{suppinfo}
\begin{equation} \label{eq:statsolpl}
\rho_{\rm eq}^m = \frac{1}{Z} \binom{2\xi}{|m|}^2\binom{2\xi+|m|}{|m|}^{-2} \simeq \frac{1}{2^{2\xi}}\binom{2\xi}{\xi+m}.
\end{equation}
The first expression is the stationary solution of \eqref{eq:LBTFull} with the Lindblad operator \eqref{eq:LBpl}, the second results if the term  $\cO(1/T)$ is dropped. Note that they approach the Gibbs state $\rho_{\rm eq}^m \sim \exp(- m^2/\xi)/Z$ as $\xi \to\infty$.

In Fig.~\ref{fig:plrot} we show the phase space dynamics of an initial superposition of two Gaussian states centered at $\alpha=0$ and $m=\pm25$. It first decoheres into a mixture which then thermalizes with the environment. The final state is given by Eq.~\eqref{eq:statsolpl}, which is well approximated by the Gibbs state.

\textit{Conclusion ---} In summary, the master equation \eqref{eq:asi} established in this Letter applies to any rotating quantum object subject to linear friction and diffusion. The associated diffusion and friction tensors, which can be determined either by a dedicated measurement or by a microscopic calculation, serve to fully characterize the effect of the thermal environment.  We found it instrumental to use a coordinate independent, tensorial formulation, rather than a specific parametrization of the rotation group and its generators. It reflects that the periodic and compact space of orientations cannot be linearized, precluding the use of standard quantum Brownian motion. Potential applications of rotational quantum Brownian motion range from ultracold chemistry with aligned molecules via torsional dynamics in molecular biophysics  and optomechanics of levitated particles to quantum rotor heat engines.

\textit{Acknowledgments---} B.A.S. and B.S. contributed equally to the manuscript. We thank S. Nimmrichter for stimulating discussions. This work was supported by the DFG (HO 2318/6-1).

\appendix
\onecolumngrid
\section{Ehrenfest equations of motion}

Denoting by ${\bf n}_i(\Omega)$ the $i$-th principal axis of the rotor so that ${\bf n}_k(\Omega) \cdot \rI(\Omega) {\bf n}_j(\Omega) = I_k \delta_{kj}$, the components of the angular momentum operator in the body fixed frame are given by $\widetilde{\op{J}}_k = {\bf n}_k \cdot \oJ = \oJ \cdot {\bf n}_k$ and in the space fixed frame by $\op{J}_k = {\bf e}_k \cdot \oJ$. They obey the commutation relation $[\op{J}_j,\op{J}_k] = i \eps_{jk\ell} \op{J}_\ell$, $[\widetilde{\op{J}}_j,\widetilde{\op{J}}_k] = -i \eps_{jk\ell} \widetilde{\op{J}}_\ell$ and $[\op{J}_j,\widetilde{\op{J}}_k] = 0$. Their commutation relations with the rotation matrix $\rR(\oOmega)$ can be expressed as
\begin{align} \label{eq:commrel}
[\op{J}_k, \rR(\oOmega)] & = \frac{\hbar}{i} {\bf e}_k \times \rR(\oOmega)  \\
[\widetilde{\op{J}}_k, \rR(\oOmega) ] & = \frac{\hbar}{i} {\bf n}_k(\oOmega) \times \rR(\oOmega).
\end{align}
Using these commutators repeatedly one obtains~(2) from the master equations~(8) and~(10).

For illustration, the dynamics of the first moment of the angular momentum operator due to (11) is
\begin{eqnarray}
\partial_t \langle \op{J}_k \rangle  & = & \frac{2 D}{\hbar^2} \left \langle {\bf m}(\oOmega) \cdot \op{J}_k {\bf m}(\oOmega)  - \op{J}_k \right \rangle - \frac{i \iGamma}{2 \hbar}  \left \langle {\bf m}(\oOmega) \cdot \op{J}_k {\bf m}(\oOmega) \times \oJ  +\oJ \times {\bf m}(\oOmega)  \cdot \op{J}_k  {\bf m}(\oOmega) \right \rangle  + \cO \left ( \frac{\hbar^2}{k_{\rm B} T I} \right ).
\end{eqnarray}
Using Eq.~\eqref{eq:commrel} with ${\bf m}(\oOmega) = \rR(\oOmega) {\bf e}_z$, the first term vanishes and the second evaluates to $ - \iGamma \langle \op{J}_k \rangle$, in accordance with (2). The calculation of the second moments follows the same lines.

\section{Linear and planar rotor thermal state}

In order to determine the stationary state of the linear rotor we consider (10) in the angular momentum eigenbasis $\ket{\ell m}$ and evaluate the matrix elements $M_{\ell m \ell' m'}^{\ell'' m''}$ defined via
\begin{equation}
\matel{\ell m}{\cD \rho_{\rm eq}}{\ell'm'} = \sum_{\ell'' = 0}^\infty \sum_{m'' = -\ell''}^{\ell''} \rho_{\rm eq}^{\ell'' m''} M_{\ell m \ell' m'}^{\ell'' m''}. 
\end{equation}
Here we used that $\rho_{\rm eq}$ is diagonal in the angular momentum basis. The matrix elements $M_{\ell m \ell' m'}^{\ell'' m''}$ can be computed by using the properties of spherical harmonics,
\begin{subequations}
\begin{align}
\oJ_1 \ket{\ell m} & = \frac{\hbar}{2} \left ( c_+ \ket{\ell m+1} + c_- \ket{\ell m-1} \right ), \\
\oJ_2 \ket{\ell m} & = \frac{\hbar}{2i} \left ( c_+ \ket{\ell m+1} - c_- \ket{\ell m-1} \right ), \\
\oJ_3 \ket{\ell m} & = \hbar m \ket{\ell m},
\end{align}
\end{subequations}
with $c_{\pm} = \sqrt{ \ell (\ell + 1) - m (m \pm 1)}$, as well as the representation of matrix elements in terms of Wigner $3$-j symbols,
\begin{align}
\matel{\ell m}{Y_{\ell'',m''}(\upbeta,\upalpha)}{\ell' m'}  = \sqrt{\frac{(2 \ell +1)(2 \ell' +1)(2 \ell'' +1)}{4 \pi}} 
\begin{pmatrix}
\ell & \ell' & \ell'' \\
0 & 0 & 0
\end{pmatrix}
\begin{pmatrix}
\ell & \ell' & \ell'' \\
-m & m' & m''
\end{pmatrix}.
\end{align}
The latter vanishes unless $m - m' - m'' = 0$ and $\ell + \ell' + \ell''$ is even, providing selection rules for the computation of the matrix elements.
These selection rules imply that the off-diagonal elements of $\matel{\ell m }{\cD\rho_{\rm eq}}{\ell'm'}$ vanish so that one has for all $\ell$, $m$
\begin{equation}
\sum_{\ell''=0}^\infty \sum_{m'' = -\ell''}^{\ell''}\rho_{\rm eq}^{\ell'' m''} M_{\ell m \ell m}^{\ell'' m''} = 0.
\end{equation}
Only a finite number of terms $\rho_{\rm eq}^{\ell'' m''}$ are coupled due to the selection rules. Starting with the equation for $\ell = 0$ and $m = 0$  one can construct the solution iteratively, arriving at Eq.~(13).

The same procedure can be used to calculate the stationary solution of the planar rotor. However, in this case one needs only the matrix elements
\begin{equation}
\matel{m}{\cos \upalpha}{m'} = \frac{1}{2} \left ( \delta_{m m+1} + \delta_{m m-1} \right ),
\end{equation}
along with $\op{p}_\alpha \ket{m} = \hbar m \ket{m}$. Again, this yields a set of equations that can be solved by iteration starting from $m=0$.

\section{Thermalization of asymmetric rotors}

We show that the Gibbs state of the asymmetric rotor is a stationary solution of (8)  for large temperatures. Note that the limit of large temperatures, $\hbar^2 / k_{\rm B} T I_{\rm min} \to 0$ with $I_{\rm min}$ the smallest moment of inertia,  is equivalent to the semiclassical limit.

We first define the transformation
\begin{equation} \label{eq:fdef}
F(\textsf{\textbf{A}}_k) = e^{-\oH/k_{\rm B} T} \textsf{\textbf{A}}_k e^{\oH/k_{\rm B} T} = \sum_{n = 0}^\infty \frac{(-k_{\rm B} T)^{-n}}{n!} \left [ \oH, \textsf{\textbf{A}}_k \right ]_n,
\end{equation}
where $[\op{A}, \op{B}]_n = [\op{A},[\op{A},\ldots,[\op{A},\op{B}]\ldots]]$ denotes the $n$-fold commutator. Note that $F(\textsf{\textbf{A}}_k \cdot \textsf{\textbf{A}}_\ell) = F(\textsf{\textbf{A}}_k) \cdot F(\textsf{\textbf{A}}_\ell)$ and $F(\textsf{\textbf{A}}_k^\dagger) \neq F(\textsf{\textbf{A}}_k)^\dagger$. With this mapping each summand of the dissipator (8) acting on the Gibbs state can be rewritten as
\begin{align} \label{eq:dk}
\cD_k \frac{e^{-\oH /k_{\rm B} T}}{Z} = & \frac{2 \widetilde{D}_k}{\hbar^2} \left ( \textsf{\textbf{A}}_k \cdot \frac{e^{-\oH /k_{\rm B} T}}{Z} \textsf{\textbf{A}}_k^\dagger - \frac{1}{2} \textsf{\textbf{A}}_k^\dagger \cdot \textsf{\textbf{A}}_k \frac{e^{-\oH /k_{\rm B} T}}{Z} - \frac{1}{2} \frac{e^{-\oH /k_{\rm B} T}}{Z} \textsf{\textbf{A}}_k^\dagger \cdot \textsf{\textbf{A}}_k \right ) \nonumber \\
 = & \frac{2 \widetilde{D}_k}{\hbar^2}\left [ \textsf{\textbf{A}}_k \cdot F(\textsf{\textbf{A}}_k^\dagger)- \frac{1}{2} \textsf{\textbf{A}}_k^\dagger \cdot \textsf{\textbf{A}}_k - \frac{1}{2} F(\textsf{\textbf{A}}_k^\dagger \cdot \textsf{\textbf{A}}_k)  \right ] \frac{e^{-\oH /k_{\rm B} T}}{Z}.
\end{align}
Inserting the expansion \eqref{eq:fdef} into \eqref{eq:dk} and sorting the terms in the square brackets in orders of $1/T$ shows that the zeroth and first order term vanish and, taking the temperature-dependence of the prefactor into account, the remainder decreases at least as $1/T$.

\section{Fokker-Planck equation of rigidly connected classical particles}

We consider $N$ point particles of mass $m_n$, position ${\bf r}_n$ and momentum ${\bf p}_n$, in an environment of temperature $T$. Denoting the friction and diffusion constant of the $n$-th particle by $\gamma_n$ and $D_n = k_{\rm B} T m_n \gamma_n$, respectively, the Fokker-Planck equation for the total phase space distribution function $f_t({\bf r}_1,\ldots,{\bf r}_N,{\bf p}_1,\ldots,{\bf p}_N)$ reads as
\begin{equation} \label{eq:fppp}
\partial_t^{\rm nc}f_t = \sum_{n = 1}^N \gamma_n  \left [  \nabla_{{\bf p}_n} \cdot \left ({\bf p}_n f_t \right ) + k_{\rm B} T m_n \nabla_{{\bf p}_n}^2 f_t \right ].
\end{equation}
This assumes that the diffusion process is isotropic.

We now invoke that the particles are rigidly connected and that their center-of-mass is fixed at the origin, so that the positions ${\bf r}_n$ are determined by the rotation matrix, ${\bf r}_n = \rR(\Omega){\bf r}_n^{(0)}$. One thus obtains for the momenta ${\bf p}_n = m_n \rI^{-1}(\Omega){\bf J} \times {\bf r}_n$ with ${\bf J} = \sum_n {\bf r}_n \times {\bf p}_n$. Exploiting that
\begin{equation}
\nabla_{{\bf p}_n} = (\nabla_{{\bf p}_n}\otimes {\bf J})\nabla_{{\bf J}} = - {\bf r}_n \times \nabla_{\bf J}
\end{equation}
yields from \eqref{eq:fppp} the rotational Fokker-Planck equation (1) with the rigid rotor distribution $h_t(\Omega,{\bf J})$. The corresponding rotational diffusion tensor can thus be identified as 
\begin{equation} \label{eq:difftens2}
\rD(\Omega) = k_{\rm B} T \sum_{n = 1}^N m_n \gamma_n \left ( r_n^2 \un - {\bf r}_n \otimes {\bf r}_n \right ).
\end{equation}
It is related to the friction tensor by $\rD(\Omega) = k_{\rm B} T \rGamma(\Omega) \rI(\Omega)$.

Note that the eigenvalues of the rotational diffusion tensor \eqref{eq:difftens2} fulfill the inequality $D_i + D_j \geq D_k$ for $(i,j,k)$ permutations of $(1,2,3)$, as can be seen from tracing over \eqref{eq:difftens2} and deducing that
\begin{equation}
\sum_{n = 1}^N m_n \gamma_n {\bf r}_n \otimes {\bf r}_n =  \frac{1}{2} \mathrm{\Tr}[ \rD(\Omega)] \un - \rD(\Omega) > 0.
\end{equation}

This constraint for the possible values of the diffusion coefficients can be relaxed by allowing for directed diffusion in Eq. \eqref{eq:fppp}. Specifically, replacing the second derivatives $\nabla_{{\bf p}_n}^2$ in the last term by $({\bf n}_n \cdot \nabla_{{\bf p}_n})^2$, so that the (particle- and orientation-dependent) unit vectors ${\bf n}_n$ define the direction of diffusion, results in the same Fokker-Planck equation (1) but with the diffusion tensor
\begin{equation}
 \rD(\Omega) = k_{\rm B} T \sum_{n = 1}^N \gamma_n m_n \left ( {\bf n}_n \times {\bf r}_n \right ) \otimes \left ( {\bf n}_n \times {\bf r}_n \right )
\end{equation}
and the corresponding friction tensor. 
Its eigenvalues can take arbitrary, positive values, depending on the $m_n$, $\gamma_n$, ${\bf n}_n$ and ${\bf r}_n$. 

\section{Inversion symmetric particles} 

The master equation (10) presupposes that the particle-bath interaction is isotropic. An inversion-symmetric particle prepared in a coherent superposition of the opposite orientations ${\bf m}(\Omega)$ and $-{\bf m}(\Omega)$ is predicted to decohere because the localization rate (12) is not zero, even if these orientations are indistinguishable by the environment. Since this symmetry enters  only on the quantum level it must not affect the semiclassical limit.

The dissipator for inversion-symmetric particles can be obtained by generalizing the microscopic derivation of inversion symmetric angular momentum diffusion [Papendell \emph{et al.}, New J. Phys. {\bf 19}, 122001 (2017)]. The Lindblad operators must then be quadratic in the particle orientation in order to preserve inversion symmetry. This yields
\begin{subequations} \label{eq:dissnLB2}
\begin{equation} \label{eq:LBfull2}
\cD \rho = \frac{D}{\hbar^2} \mathrm{Tr} \left [ \mathrm{\sf B} \rho \mathrm{\sf B}^\dagger - \frac{1}{2} \left \{ \mathrm{\sf B}^\dagger \mathrm{\sf B}, \rho \right \} \right ],
\end{equation}
where $\mathrm{Tr}(\cdot)$ denotes the matrix trace (not to be confused with the operator trace) and the tensor Lindblad operators are
\begin{equation} \label{eq:LB2}
\mathrm{\sf B} = {\bf m}(\oOmega) \otimes {\bf m}(\oOmega) - \frac{i \hbar}{2 k_{\rm B} T} {\bf m}(\oOmega) \otimes {\bf m}(\oOmega) \times \rI^{-1}(\oOmega) \oJ.
\end{equation}
\end{subequations}
While the first term appears already in the article by Papendell \emph{et al.}, the second results from quantizing the time derivative $\partial_t[{\bf m}(\Omega)\otimes {\bf m}(\Omega)]$. The latter can be expressed as  ${\bf m}(\Omega) \otimes {\bf m}(\Omega) \times \rI^{-1}(\Omega) {\bf J}$ because of the matrix trace in \eqref{eq:LBfull2} without affecting diffusion and friction.

The dissipator \eqref{eq:dissnLB2} preserves inversion symmetry and implies the moment equations of motion (2) as well as the thermalization (3) and (4). In addition, it also leads to the Fokker-Planck equation (1). The $T$-independent contribution of \eqref{eq:LB2} depends only on the orientation operator and thus leads to orientational decoherence and angular momentum diffusion. The corresponding decoherence rate
\begin{equation}
F(\Omega,\Omega') = \frac{k_{\rm B} T \iGamma I}{\hbar^2} \vert {\bf m}(\Omega)\times {\bf m}(\Omega')\vert^2,
\end{equation}
vanishes not only for $\Omega = \Omega'$ but also for a superpositions between opposite orientations.
The quantum phase space dynamics of the inversion-symmetric planar rotor can be obtained from Eq.~(10) by replacing $\iGamma$ by $\iGamma/2$, $D$ by $D/4$ and $m \pm 1$ by $m \pm 2$.

\twocolumngrid
\end{document}